\documentclass[prb,twocolumn,preprintnumbers,amsmath,amssymb]{revtex4}

\usepackage{graphicx}
\usepackage{dcolumn}
\usepackage{bm}

\begin{document}

\title{Impurity Effect on the In-plane Penetration Depth of the Organic 
Superconductors $\kappa$-(BEDT-TTF)$_2X$ ($X$ = Cu(NCS)$_2$ and 
Cu[N(CN)$_2$]Br)}

\author{N. Yoneyama, A. Higashihara, T. Sasaki}
\affiliation{Institute for Materials Research, Tohoku University, Sendai
980-8577, Japan}
\author{T. Nojima}
\affiliation{Center for Low Temperature Science, Tohoku University, Sendai 
980-8577, Japan}
\author{N. Kobayashi}
\affiliation{Institute for Materials Research, Tohoku University, Sendai
980-8577, Japan\\
Center for Low Temperature Science, Tohoku University, Sendai 
980-8577, Japan}

\date{\today}

\begin{abstract}
We report the in-plane penetration depth $\lambda_{\parallel}$ of
single crystals $\kappa$-(BEDT-TTF)$_2X$ ($X=$ Cu(NCS)$_2$ and 
Cu[N(CN)$_2$]Br) by means of the reversible magnetization measurements
under the control of cooling-rate.
In $X$ = Cu(NCS)$_2$, $\lambda_{\parallel}(0)$ as 
an extrapolation toward $T$ = 0 K does not change by the 
cooling-rate within the experimental accuracy, while $T_{\textrm{c}}$ is slightly 
reduced.
On the other hand, in $X$ = Cu[N(CN)$_2$]Br, $\lambda_{\parallel}(0)$ 
indicates a distinct increase by cooling faster.
The different behavior of $\lambda_{\parallel}(0)$ on cooling-rate 
between the two salts is quantitatively explained in terms of the 
local-clean approximation (London model), considering that 
the former salt belongs to the very clean system and the later the moderate 
clean one.
The good agreement with this model demonstrates that disorders of 
ethylene-group in BEDT-TTF introduced by cooling faster increase the 
electron(quasiparticle)-scattering, resulting in shorter mean free 
path.
\end{abstract}

\maketitle

\section{Introduction}
The organic superconductors $\kappa$-(BEDT-TTF)$_2X$ ($X$ = Cu(NCS)$_2$ and 
Cu[N(CN)$_2$]Br) belong to a family of charge transfer salts, where 
BEDT-TTF is bis(ethylenedithio)-tetrathiafulvalene.
The crystal structure consists of alternating layers of conducting 
BEDT-TTF and insulating $X$ anions.\cite{Ishiguro}
Consequently, these systems are characterized by highly anisotropic 
properties. 
High purity in samples makes it possible to observe 
the quantum oscillation effects,\cite{853,159,851,667,1025,680,838,278} 
demonstrating the quasi-two-dimensional electronic states. 
In the superconducting state, these salts are described in terms of 
type-II superconductor with long in-plane penetration depth 
$\lambda_{\parallel}$ and short in-plane coherence length 
$\xi_{\parallel}$ as described below.

\begin{figure}
\begin{center}
\includegraphics[clip]{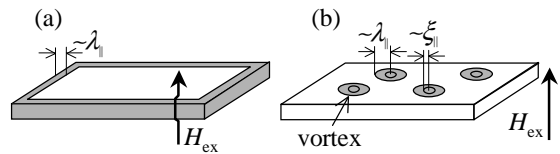}
 \caption{\label{Fig:schematicview} Schematic view of the in-plane 
 penetration depth in (a) the shielding state and (b) the mixed state.
 Bold arrows indicate the magnetic-field perpendicular to the conduction plane.
 The gray-shaded regions indicate the penetrating area of the magnetic-field.
 }
\end{center}
\end{figure}

The in-plane penetration depth is one of the fundamental parameters of 
superconductivity, because the temperature-dependence of 
$\lambda_{\parallel}$ at low temperatures is reflected by gap 
excitation in the superconducting state, giving information on the gap 
symmetry, i.e., full-gap or node-gap, etc.
Thus there have been a large number of reports on $\lambda_{\parallel}$ as a 
function of 
temperature.\cite{994,141,108,793,153,995,181,177,163,197,185,20,785,180,220,1026} 
However, prior to such a delicate discussion, it should be noticed that
the wide distribution of the absolute values of $\lambda_{\parallel}(0)$
has been reported, taking into account the following inconsistency in 
$\lambda_{\parallel}(0)$ between experimental methods. 
The values of $\lambda_{\parallel}(0)$ in the 
literatures\cite{994,141,108,793,153,995,181,177,163,197,185,20,785,180,220,1026}
are collected in Tables \ref{t1} and \ref{t2}.
All the measurements have been carried out at the same configuration of 
magnetic-field $H$ perpendicular to the conduction plane.
Nevertheless one can find that they are classified into two groups; 
one is with $\lambda_{\parallel}(0)$ longer than $\sim$ 1 $\mu$m (surface 
impedance,\cite{994,141} ac inductance,\cite{108} and ac 
susceptibility\cite{20,785}) and the other with shorter 
$\lambda_{\parallel}(0)$ ($\mu$SR,\cite{995,177,163} 
magnetization,\cite{197,180,220} and decoration method\cite{185}). 
The former measurements are performed in magnetic-field lower than 
$H_{\textrm{c1}}$.
In this situation, shielding current flows around the sample surface 
to exclude $H$ (shielding states), and then $\lambda_{\parallel}$ is 
obtained as the 
penetrating length of $H$ at the sample edge (Fig. \ref{Fig:schematicview}(a)).
On the other hand, the latter measurements are carried out in 
magnetic-field much larger than $H_{\textrm{c1}}$, where $H$ penetrates into the 
sample as vortices (mixed states). 
In this case, $\lambda_{\parallel}$ is given as a decay length of the
magnetic-field from the center of the vortex (Fig. \ref{Fig:schematicview}(b)).
As shown in Tables \ref{t1} and \ref{t2}, the former group 
seems to overestimate $\lambda_{\parallel}(0)$ compared to the latter 
one, because the former methods may be sensitive to the surface states.
Up to now, only little effort has been spent on this controversial point.

In the superconducting state of $X$ = Cu(NCS)$_2$, hereafter abbreviated 
as the Cu(NCS)$_2$ salt, the local and clean limit is adequate to the 
systems; $\lambda_{\parallel} > l_{\parallel} > \xi_{\parallel}$,
where the intermediate in-plane mean free path $l_{\parallel}=100-240$ 
nm\cite{853,159,851,667,1025,estimate_l} and the short $\xi_{\parallel}=
3.1-7$ nm.\cite{177,141,138}  
Compared with this salt, $l_{\parallel}$ of 
$X$ = Cu[N(CN)$_2$]Br (the Cu[N(CN)$_2$]Br salt) is remarkably short: 
$l_{\parallel}=26-38$ nm\cite{680,838,278}, while 
$\xi_{\parallel}$ ($=2.4-3.7$ nm\cite{141,138}) is almost comparable to 
the Cu(NCS)$_2$ salt.
On the basis of such the local-clean approximation (London model),\cite{Tinkham} 
 $\lambda_{\parallel}(0)$ is described as
\begin{equation}
\lambda_{\parallel}(0)=\lambda_{\textrm{L}}(0)(1+\xi_0/l_{\parallel})^{0.5},\label{Eq:London}
\end{equation}
where $\lambda_{\textrm{L}}(0)$ is the London penetration depth for a pure sample and 
$\xi_0$ is the coherence length. 
In eq. (\ref{Eq:London}), when $l_{\parallel}$ becomes short with 
increasing impurity scattering, $\lambda_{\parallel}(0)$ becomes longer than $\lambda_{\textrm{L}}(0)$.

A possible impurity of the present materials is originated from a  
feature of molecular-based compounds, because an internal degree of 
freedom in the donor molecule can give a positional disorder.
At room temperature, conformation of the terminal ethylene groups in
BEDT-TTF is thermally excited.\cite{231}  
Such conformational disorders (ethylene-disorders) are frozen by
cooling very fast (quenching) and may affect the scattering of carriers 
at low temperatures.
Especially in the Cu[N(CN)$_2$]Br salt, as mentioned above, 
$l_{\parallel}$ is shorter than that of the Cu(NCS)$_2$ salt, and so the 
electronic system of the Cu[N(CN)$_2$]Br salt will be slightly dirty. 
In addition, several investigations have been reported in the 
Cu[N(CN)$_2$]Br salt under the control of cooling-rate.
The residual resistivity increases by cooling faster and simultaneously
the superconducting transition temperature $T_{\textrm{c}}$ decreases.\cite{146}
In a Shubnikov-de Haas (SdH) oscillation measurement,\cite{278} the 
oscillation amplitude is suppressed with increasing disorders.
These results imply that the ethylene-disorders play an important role 
on the electron (or quasiparticle)-scattering as an impurity origin.

Aburto \textit{et al}.\cite{220} have reported a slight increase of 
$\lambda_{\parallel}(0)$ by cooling fast (a magnetization measurement in 
the mixed state). 
On the other hand, an ac susceptibility measurement\cite{785} (in the 
shielding state) provides an extremely large increase of  
$\lambda_{\parallel}(0)$ by quenching, up to $\sim 100\ \mu$m.
Motivated by this inconsistency in the behavior of 
$\lambda_{\parallel}(0)$ between the experimental methods, we present in 
this paper a quantitative explanation for the increase of 
$\lambda_{\parallel}(0)$ by the impurity effect originated from ethylene-disorders.

We evaluate $\lambda_{\parallel}(0)$ from dc reversible magnetization 
measurements in the mixed state under the control of cooling-rate.
This method gives a reliable estimate of the penetration depth in bulk samples, 
because a homogeneous magnetic-field penetration appears around vortices 
free from pinning.
This investigation reveals that a distinct difference appears between the two 
salts on the behavior of the cooling-rate dependence of $\lambda_{\parallel}(0)$.
In the Cu(NCS)$_2$ salt, $\lambda_{\parallel}(0)$ is not changed by 
quenching within the experimental accuracy, while $T_{\textrm{c}}$ decreases slightly.
In contrast, in the Cu[N(CN)$_2$]Br salt, a significant change in 
$\lambda_{\parallel}(0)$ is observed with increasing cooling-rate. 
We demonstrate that this behavior is well described using the local-clean 
limit approximation in eq. (\ref{Eq:London}).
The ethylene-disorders introduced by cooling faster will increase 
the electron (quasiparticle)-scattering, resulting in shorter mean free path.
A preliminary report is given in Ref. \onlinecite{yone2}.

\begin{table}[tb]
\caption{A list of in-plane penetration depths in $\kappa$-(BEDT-TTF)$_2$Cu(NCS)$_2$.}
\label{t1}
\begin{tabular}{cccc}
\hline
magnetic-field & experiments & $\lambda_{\parallel}(0)$ ($\mu$m) & Ref. \\
\hline
$H < H_{\textrm{c1}}$ & surface impedance & 2 & \onlinecite{994} \\
 & surface impedance & 1.4 & \onlinecite{141} \\
 & ac inductance & 1.8  & \onlinecite{108} \\
 & ac susceptibility & -- & \onlinecite{793,153} \\
\cline{2-4}
$H > H_{\textrm{c1}}$ & $\mu$SR & 0.98 & \onlinecite{995} \\
 & $\mu$SR & 0.768 & \onlinecite{177} \\
 & $\mu$SR & 0.43 & \onlinecite{163} \\
 & decoration & 0.4 & \onlinecite{185} \\
 & magnetization & 0.535 & \onlinecite{197} \\
 & magnetization & 0.43 & this work \\
\hline
\end{tabular}
\end{table}

\begin{table}[tb]
\caption{A list of in-plane penetration depths in $\kappa$-(BEDT-TTF)$_2$Cu[N(CN)$_2$]Br.}
\label{t2}
\begin{tabular}{cccc}
\hline
magnetic-field & experiments & $\lambda_{\parallel}(0)$ ($\mu$m) & Ref. \\
\hline
$H < H_{\textrm{c1}}$ & surface impedance & 1.5 & \onlinecite{141} \\
 & ac inductance & 3.2 & \onlinecite{108} \\
 & ac susceptibility & $\sim$ 1 & \onlinecite{20} \\
 & ac susceptibility & 1.5(R)--100(Q)\footnote{R: $\sim$0.2 K/min, Q: $\sim$300 K/min.} & \onlinecite{785} \\
 & ac susceptibility & 1.1(A)--24(Q)\footnote{A: annealed at 78 -- 100 K, Q: $\sim$300 K/min.} & \onlinecite{785} \\
\cline{2-4}
$H > H_{\textrm{c1}}$ & $\mu$SR & -- & \onlinecite{181} \\
 & decoration & -- & \onlinecite{1026} \\
 & magnetization & 0.65 & \onlinecite{180} \\
 & magnetization & 0.58(S)--0.64(R)\footnote{S: 1.4 K/min, R: 20 K/min.} & \onlinecite{220} \\
 & magnetization & 0.57(S)--0.69(Q)\footnote{S: slow-cooled, Q: quenched.}  & this work \\
\hline
\end{tabular}
\end{table}

\section{Experimental}
Single crystals were grown by a standard electrochemical technique.
The dimensions of samples were $2.2 \times 1.1 \times  0.40$ cm$^3$ 
(the Cu(NCS)$_2$ salt) and $1.1 \times 1.0 \times 0.13 \mathrm{mm}^3$ 
(Cu[N(CN)$_2$]Br).
The samples were cooled with a rate of $\sim$100 K/min from room 
temperature to 15 K,  making ``quenched'' state.
Following the measurements in the quenched state, two cooling procedures 
were carried out in series with 15 K/min (rapid-cooled) and 0.2 K/min 
(slow-cooled) in each cooling process after warmed up to 100 K.
In the different cooling states, a series of the magnetization measurements
was performed using a SQUID magnetometer (Quantum Design, MPMS-5 and 
MPMS-7).
The magnetic-field was applied perpendicular to the conduction plane.
The similar experiments for other crystals indicated good 
reproducibility with the data presented here for both salts.

For type-II superconductor in the magnetic-field regime ($H_{\textrm{c1}} <) 
H_{\textrm{irr}} < H \ll H_{\textrm{c2}}$,
the London model\cite{deGennes,891} describes the magnetization $M$ as a 
function of $H$ as \begin{equation}
-4\pi M=(\alpha\phi_0/8\pi \lambda_{\parallel}^2)\ln(H_{\textrm{c2}}\beta/H),
\label{Eq:MH}
\end{equation}
where $H_{\textrm{irr}}$ is the irreversibility field of the magnetizations,
$\phi_0$ the magnetic flux quantum, and $\beta$ a constant of order unity.
While a correction factor $\alpha$ is unity for the conventional London model, 
Hao and Clem\cite{891} have corrected the vortex-core contribution neglected in 
the London model, which provided $\alpha$ = 0.70 in the present 
magnetic-field range.
According to eq. (\ref{Eq:MH}), $\lambda_{\parallel}$ was estimated 
from the slope of the linear dependence in the $M$ vs $\ln H$ plot, 
where the slope is $\alpha\phi_0/32\pi^2\lambda_{\parallel}^2$. 

\section{Results}
\begin{figure}
\begin{center}
\includegraphics[clip]{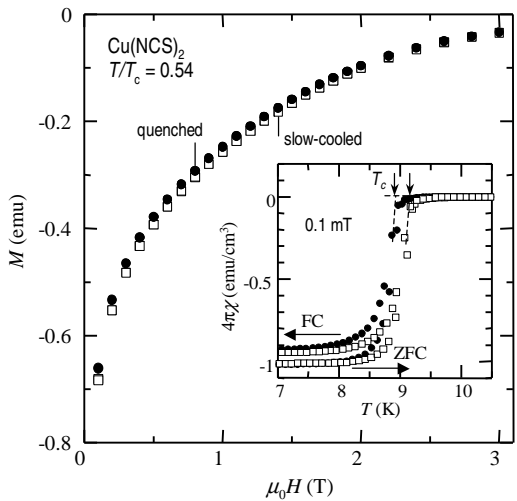}
 \caption{\label{Fig:MH1}Magnetization curves of 
$\kappa$-(BEDT-TTF)$_2$Cu(NCS)$_2$ at $T/T_{\textrm{c}}$ = 0.54 for the 
slow-cooled (open squares, $T/T_{\textrm{c}}$ = 5.0 K/9.2 K) 
 and quenched states (filled circles, $T/T_{\textrm{c}}$ = 4.8 K/8.9 K). 
 The inset shows the temperature  dependence of the magnetic 
 susceptibilities at 0.1 mT under zero-field-cooling (ZFC) and 
 field-cooling (FC) conditions. 
 }
\end{center}
 \end{figure}

\begin{figure}
\begin{center}
\includegraphics[clip]{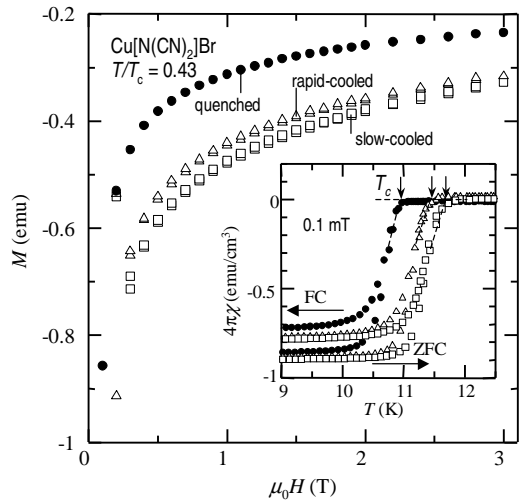}
 \caption{\label{Fig:MH2}Magnetization curves of 
 $\kappa$-(BEDT-TTF)$_2$Cu[N(CN)$_2$]Br at $T/T_{\textrm{c}}$ = 0.43 for the 
 slow-cooled (open squares, $T/T_{\textrm{c}}$ = 5.0 K/11.7 K), 
rapid-cooled (open triangles, $T/T_{\textrm{c}}$ = 4.9 K/11.4 K), and quenched 
states (filled circles, $T/T_{\textrm{c}}$ = 4.7 K/10.9 K). 
The inset shows the temperature dependence of the magnetic 
susceptibilities at 0.1 mT under zero-field-cooling (ZFC) and 
field-cooling (FC) conditions.
}
\end{center}
\end{figure}

\begin{figure}
\begin{center}
\includegraphics[clip]{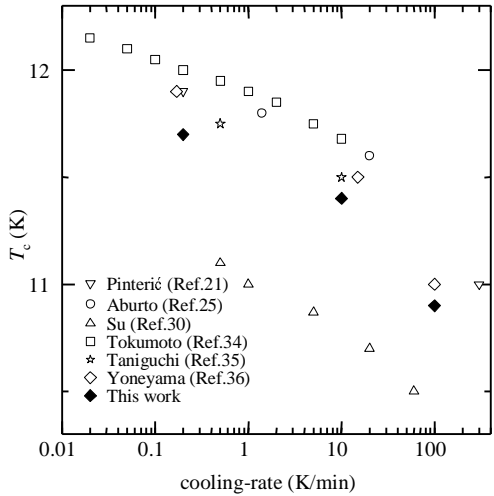}
 \caption{\label{Fig:Tc} Cooling-rate dependence of $T_{\textrm{c}}$ in 
 $\kappa$-(BEDT-TTF)$_2$Cu[N(CN)$_2$]Br. The data of the present work
are displayed with filled diamonds. Open symbols are the data taken from the 
literatures.\cite{146,227,220,785,203,yone}
}
\end{center}
\end{figure}

The insets of Figs. \ref{Fig:MH1} and \ref{Fig:MH2} show the 
magnetic susceptibility of $X$ = Cu(NCS)$_2$ and 
Cu[N(CN)$_2$]Br at 0.1 mT, respectively.
After subtracting the contribution of the core diamagnetization, 
the demagnetization factor was corrected using an ellipsoidal 
approximation.
In both salts, $T_{\textrm{c}}$ decreases with increasing cooling-rate: 9.2 K 
(slow-cooled) and 8.9 K (quenched) for the Cu(NCS)$_2$ salt, 11.7 K 
(slow-cooled), 11.4 K (rapid-cooled), and 10.9 K (quenched) for the 
Cu[N(CN)$_2$]Br salt.
Here $T_{\textrm{c}}$'s are defined as an intercept of the extrapolated lines of the
normal and superconducting states (broken lines and arrows in the insets 
of Figs. \ref{Fig:MH1} and \ref{Fig:MH2}).
The cooling-rate dependence of $T_{\textrm{c}}$ for the Cu[N(CN)$_2$]Br salt is 
consistent with the literatures,\cite{146,227,220,785,203,yone} as 
shown in Fig. \ref{Fig:Tc}.
This plot indicates that $T_{\textrm{c}}$ changes with almost an unified slope as a 
function of cooling-rate $s$ (K/min): $|\Delta T_{\textrm{c}}/\Delta \log(s)|=
0.2-0.3$, whereas the absolute values of $T_{\textrm{c}}$ different between the
reports may come from the different definition of $T_{\textrm{c}}$.
The magnitude of the reduction in $T_{\textrm{c}}$ for the Cu(NCS)$_2$ salt is much 
smaller than that for the Cu[N(CN)$_2$]Br salt.
The susceptibilities under a zero-field-cooled condition (ZFC)
indicate almost the full Meissner volume.
The negligible cooling-rate dependence of the superconducting volume is
consistent with the reported results,\cite{203,yone}
whereas Pinteri\'{c} \textit{et al}. have reported a small suppression of the
volume by cooling faster.\cite{785}

The main panels of Figs. \ref{Fig:MH1} and \ref{Fig:MH2} show the 
magnetization curves at $T/T_{\textrm{c}}$ = 0.54 (the Cu(NCS)$_2$ salt) and 0.43 
(the Cu[N(CN)$_2$]Br salt).
The data at 5.0 K are shown for both slow-cooled states, and the 
corresponding data with the same $T/T_{\textrm{c}}$ value for the rapid-cooled and 
quenched states are estimated from linear interpolations with the data 
at 4.5 and 5.0 K. 
The irreversibility field around 5 K is about 80 mT for the Cu(NCS)$_2$ 
salt and about $200-300$ mT for the Cu[N(CN)$_2$]Br salt.
Within the experimental accuracy, no change of the reversible 
$M$ by quenching is observed in the Cu(NCS)$_2$ salt.
On the contrary, in the Cu[N(CN)$_2$]Br salt, a large difference in $M$ 
appears by changing cooling-rate.

\begin{figure}
\begin{center}
\includegraphics[clip]{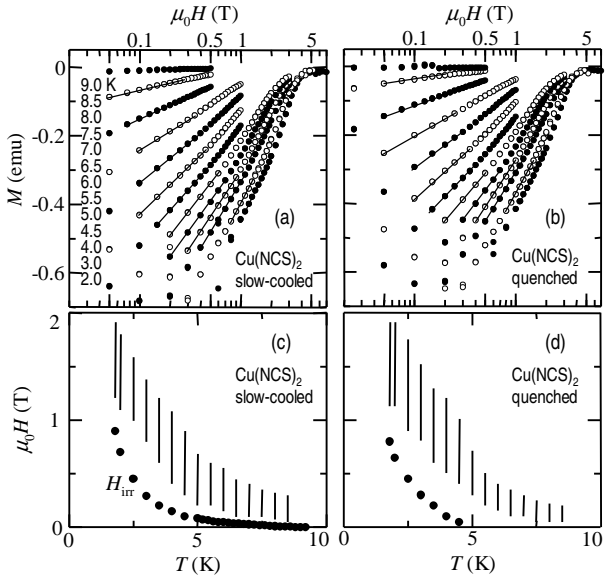}
 \caption{\label{Fig:MH3}Magnetization vs $\ln H$ plots in
 $\kappa$-(BEDT-TTF)$_2$Cu(NCS)$_2$ for (a) the slow-cooled and 
 (b) quenched states. Lower panels (c) and (d) indicate the 
 irreversibility field $H_{\textrm{irr}}$ (filled circles) and the 
 region for the linear reversible magnetization in ln$H$
 (solid vertical bars) for each cooling-rate.
}
 \end{center}
\end{figure}

\begin{figure}
\begin{center}
\includegraphics[clip]{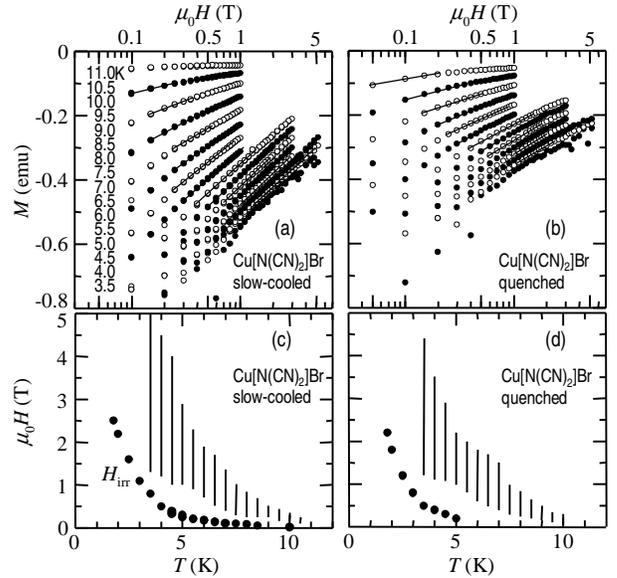}
 \caption{\label{Fig:MH4}Magnetization vs $\ln H$ plots in
 $\kappa$-(BEDT-TTF)$_2$Cu[N(CN)$_2$]Br for (a) the slow-cooled and 
 (b) quenched states. Lower panels (c) and (d) indicate the 
 irreversibility field $H_{\textrm{irr}}$ (filled circles) and the 
 region for the linear reversible magnetization in ln$H$
 (solid vertical bars) for each cooling-rate.
}
\end{center}
\end{figure}

The reversible magnetizations are depicted on a semilogarithmic 
scale in Figs. \ref{Fig:MH3}(a), \ref{Fig:MH3}(b), \ref{Fig:MH4}(a), and 
\ref{Fig:MH4}(b).
In this plot a linear-dependent region emerges (solid 
lines), except for the data of the Cu[N(CN)$_2$]Br salt below 3.0 K 
(not shown in the Figures).
The linear regions are also displayed in Figs. \ref{Fig:MH3}(c), \ref{Fig:MH3}(d),
\ref{Fig:MH4}(c), and \ref{Fig:MH4}(d) as solid vertical bars,
which satisfy the condition $H_{\textrm{irr}} < H \ll H_{\textrm{c2}}$ and thus 
guarantee the appropriate determination of the penetration depth.

\begin{figure}
\begin{center}
\includegraphics[clip]{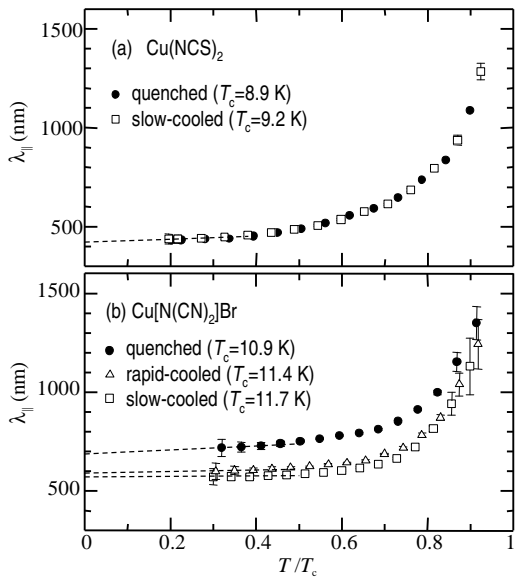}
 \caption{\label{Fig:lambda} Temperature dependence of the in-plane 
 penetration depth for $\kappa$-(BEDT-TTF)$_2X$, where $X$ = (a) 
 Cu(NCS)$_2$ and (b) Cu[N(CN)$_2$]Br. Broken lines are linear 
 extrapolations toward $T$ = 0 K, providing $\lambda_{\parallel}(0)$.
 }
\end{center}
\end{figure}

From these slopes, the penetration depths are obtained
as shown in Fig. \ref{Fig:lambda}.
In the Cu(NCS)$_2$ salt, we do not find any difference in
$\lambda_{\parallel}(T)$ at low temperatures between the two cooling-rates
within the accuracy of the measurement.
In order to estimate $\lambda_{\parallel}(0)$, we adopt
a linear extrapolation to $T$ = 0 K, giving $\lambda_{\parallel}(0)$ = 
430$\pm20$ nm (slow-cooled and quenched).
Here, it should be noted that it is difficult to obtain meaningful fits to the
present data as a function of $T$, and thus the $T$-linear dependence 
employed here is not worth discussing in detail.

In contrast to the Cu(NCS)$_2$ salt, a distinct increase of 
$\lambda_{\parallel}$ by cooling faster is observed in the Cu[N(CN)$_2$]Br salt.
Similar estimates of $\lambda_{\parallel}(0)$ by 
linear extrapolations result in  570$\pm30$ nm (slow-cooled), 
590$\pm30$ nm (rapid-cooled), and 690$\pm50$ nm (quenched).
This change of $\lambda_{\parallel}(0)$ can be also seen in the 
original $M(\ln H)$ curves shown in Fig. \ref{Fig:MH4}, where the slope 
in the slow-cooled state (Fig. \ref{Fig:MH4}(a)) is suppressed by 
quenching (Fig. \ref{Fig:MH4}(b)), while the corresponding data for the 
Cu(NCS)$_2$ salt do not show such a cooling-rate dependence of the 
slope (Figs. \ref{Fig:MH3}(a,b)). 

\section{Discussion}

First we compare the absolute values of $\lambda_{\parallel}(0)$ in
the slow-cooled state with the reported results summarized in Tables 
\ref{t1} and \ref{t2}; $\lambda_{\parallel}(0)$ obtained in the present work 
is 430$\pm20$ and 570$\pm30$ nm for the Cu(NCS)$_2$ and Cu[N(CN)$_2$]Br 
salts, respectively.
The values of $\lambda_{\parallel}(0)$ are almost comparable to the data by 
magnetization,\cite{197,180,220} $\mu$SR,\cite{995,177,163} and decoration\cite{185}
measurements, which have been carried out in the mixed states.
The other reports performed in the shielding states, i.e., surface 
impedance,\cite{994,141}  
ac susceptibility,\cite{20,785} and ac inductance\cite{108} 
methods, provide 1--2 $\mu$m, which is one order of magnitude larger 
than our data.
This remarkable difference in $\lambda_{\parallel}(0)$ between the 
experimental methods can be explained in terms of the different
''field-penetrated portion for evaluating $\lambda$'' as displayed in Fig. 
\ref{Fig:schematicview}. 
The details are discussed later.

In the slow-cooled state, the number of ethylene-disorders will be small
and thus it is fruitful to compare with the London penetration depth 
$\lambda_{\textrm{L}}(0)$.
It can be evaluated by means of the relation:
$\lambda_{\textrm{L}}(0)=(c/e)(m^*/4\pi n_{\textrm{s}}(0))^{1/2}$, where $m^*$ is the 
effective mass and $n_{\textrm{s}}(0)$ the carrier density.
For the Cu(NCS)$_2$ salt, Lang \textit{et al}.\cite{197} have estimated 
$\lambda_{\textrm{L}}(0)$ to be 410$\pm40$ nm taking account of both $\alpha$- and 
$\beta$-Fermi surfaces contributing to the superconducting state.
The values of $\lambda_{\parallel}(0)$ obtained in the present experiment is in good 
agreement with $\lambda_{\textrm{L}}(0)$. 
This demonstrates that the Cu(NCS)$_2$ salt is very clean and 
$l_{\parallel}$ is much long enough to neglect the term 
``$\xi_0/l_{\parallel}$'' in eq.(\ref{Eq:London}), resulting in 
$\lambda_{\parallel}(0)\approx \lambda_{\textrm{L}}(0)$. 
In the Cu[N(CN)$_2$]Br salt, one obtains $\lambda_{\textrm{L}}(0)=360-390$ nm 
by a similar calculation using $m^*/m_0=5.4-6.7$ estimated from 
quantum oscillation investigations\cite{278,680,838} and 
$n_{\textrm{s}}(0)=1.2\times10^{21}$ cm$^{-3}$ ($\beta$ orbital).
Thus, $\lambda_{\parallel}(0)$ is much longer than the value of 
$\lambda_{\textrm{L}}(0)$ evaluated.
This can also be understood in terms of eq. (\ref{Eq:London});
in the Cu[N(CN)$_2$]Br salt, ``$\xi_0/l_{\parallel}$'' should not be neglected even 
in the slow-cooled state because $l_{\parallel}$ is much shorter than that of the 
Cu(NCS)$_2$ salt.
Therefore the effect of $l_{\parallel}$, i.e., impurity (i.e., cooling-rate as 
mentioned below) on $\lambda_{\parallel}$ is 
more effective on the Cu[N(CN)$_2$]Br salt than the Cu(NCS)$_2$ salt.

\begin{figure}
\begin{center}
\includegraphics[clip]{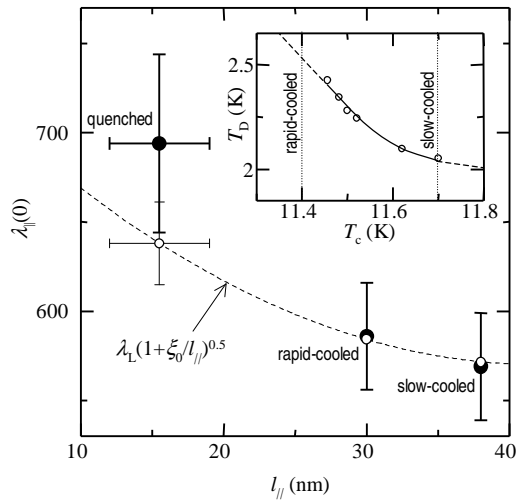}
 \caption{\label{Fig:Londonfit}In-plane penetration depth as a function 
 of the in-plane mean free path $l_{\parallel}$ for 
 $\kappa$-(BEDT-TTF)$_2$Cu[N(CN)$_2$]Br (closed circles).
 Small open circles are calculated from the London model within the 
local-clean approximation with a fitting parameter 
$\lambda_{\textrm{L}}(0)$ = 523 nm. The broken curve is a guide to the eye.
Inset shows the relationship between Dingle temperature $T_{\textrm{D}}$ and the 
superconducting transition temperature $T_{\textrm{c}}$, extracted from the 
literature.\cite{278}
 }
\end{center}
\end{figure}

Before discussing the cooling-rate effect on $\lambda_{\parallel}(0)$, 
we review the impurity effect by ethylene-disorders in the 
Cu[N(CN)$_2$]Br salt.
It will be believed that the number of the ethylene-disorders increases with
cooling faster.
Although no direct observation of the disorders has been 
succeeded in such as X-ray structural studies at low temperatures,\cite{231}
the number of the disorders has been proposed to be about 
5--20\%.\cite{170,329}
Moreover, the investigation of the SdH effect under the control of 
disorders\cite{278} implies that these disorders can work as an 
electron (quasiparticle)-scattering origin.
They have found a relationship between $T_{\textrm{c}}$ and the Dingle temperature 
$T_{\textrm{D}}$, as shown in the inset of Fig. \ref{Fig:Londonfit}.
With decreasing $T_{\textrm{c}}$, $T_{\textrm{D}}$ monotonically increases.
Generally $T_{\textrm{D}}$ reflects the sample purity, 
since the mean free path $l$ is expressed as $l=v_{\textrm{F}}\hbar/2\pi k_{\textrm{B}}T_{\textrm{D}}$,
where $v_{\textrm{F}}$ is the Fermi velocity.
Therefore, a state with larger $T_{\textrm{D}}$ corresponds to a more 
dirty system with shorter $l$.
In the following, we explain the increase of $\lambda_{\parallel}(0)$ by 
cooling faster in the Cu[N(CN)$_2$]Br salt on the basis of eq.
(\ref{Eq:London}) using the mean free path evaluated.

As shown in the inset of Fig. \ref{Fig:Londonfit}, $T_{\textrm{D}}$ is about 2.0 K 
and 2.5 K for the slow-cooled and rapid-cooled state, respectively.
From these values, $l_{\parallel}$ is obtained as
38 nm (slow-cooled) and 30 nm (rapid-cooled).
In a state with much lower $T_{\textrm{c}}$, like as being 10.9 K for the quenched state, 
the SdH oscillation amplitudes are considerably suppressed,\cite{278} 
and thus $T_{\textrm{D}}$  has not been obtained.
Then an extrapolation to $T_{\textrm{c}}$ = 10.9 K for the quenched 
state gives a very rough value of $T_{\textrm{D}}\sim 5\pm1$ K, 
corresponding to $l_{\parallel}=12-19$ nm.  
This is still larger than $\xi_{\parallel}=2.4-3.7$ nm\cite{141,138}, and 
therefore the clean-local limiting London model is suitable to apply to 
the present systems even in the quenched state.
Figure \ref{Fig:Londonfit} shows $\lambda_{\parallel}(0)$ obtained 
experimentally as a function of the estimated $l_{\parallel}$ 
(filled circles).
The corresponding values of $\lambda_{\parallel}(0)$ evaluated from eq. 
(\ref{Eq:London}) is also plotted using a fitting parameter of 
$\lambda_{\textrm{L}}(0)$ = 523 nm and values of $\xi_0$ calculated from 
the relation $\xi_0 = a \hbar 
v_{\textrm{F}}/k_{\textrm{B}}T_{\textrm{c}}$, where $a$ = 0.18 and 
$v_{\textrm{F}}=6.2\times10^6$ cm/s (open circles). 
The experimental data are in good quantitative agreement with this model.
The deviation in the quenched data may be attributed to the 
ambiguity in the too roughly estimated value of $l_{\parallel}$.
The good accordance with this model
demonstrates that the penetration depth is well described in 
terms of the clean-local approximation, and the introduction of the disorders 
provides the impurity effect on the electron (quasiparticle)-scattering, resulting in
shorter mean free path.

We here comment on the quite large increase of $\lambda_{\parallel}(0)$ 
from the experiments in the shielding states by quenching. 
From the present investigation (in the mixed state), we obtain 
a ratio of the penetration depth in the faster cooling state (quenched 
or rapid-cooled) ($\lambda_{\parallel}^{\textrm{q}}(0)$) to that in the slow-cooled one
($\lambda_{\parallel}^{\textrm{s}}(0)$) as $\lambda_{\parallel}^{\textrm{q}}(0)/
\lambda_{\parallel}^{\textrm{s}}(0) \sim 1.03-1.21$. 
Aburto \textit{et al}.\cite{220} have reported the similar magnitude of 
the ratio by means of the magnetization measurements at high 
magnetic-fields, 
$\lambda_{\parallel}^{\textrm{q}}(0)/ \lambda_{\parallel}^{\textrm{s}}(0)\sim 1.1$,
where $\lambda_{\parallel}(0)$ = 640 nm (faster cooling of 20 K/min, 
$T_{\textrm{c}}$ = 11.6 K) and 580 nm (slower cooling of 1.4 K/min, 
$T_{\textrm{c}}$ = 11.8 K).
As discussed above, this change of the absolute values is
quantitatively consistent with the London model, corresponding to 
an increase of about 10--20\% by quenching.
In other words, the disorders do not lead
to more than 10--20\% increase of $\lambda_{\parallel}(0)$ at most.
In contrast, the measurements in the shielding state 
not only provide much larger $\lambda_{\parallel}(0)$ in the 
slow-cooled state, but also result in a very large 
increase of $\lambda_{\parallel}(0)$ by quenching (ac susceptibility\cite{785}),
where the ratio $\lambda_{\parallel}^{\textrm{q}}(0)/ \lambda_{\parallel}^{\textrm{s}}(0)$ is 
estimated to be about 22--67. 
This increase of $\lambda_{\parallel}(0)$ by quenching is two orders of 
magnitude larger than our results, and also the value expected from the 
London model for bulk samples. 
On this discrepancy, we point out that these anomalously large 
values of $\lambda_{\parallel}(0)$ observed in the shielding state are
attributed to a surface effect of samples.
In the experimental techniques which observe shielding volumes at very 
low magnetic-field below $H_{\textrm{c1}}$ (like ac susceptibility),
the penetration depth is obtained from the difference between the shielding 
volume and the sample volume, assuming the magnetic-field is penetrating 
on the crystalline edge (Fig. \ref{Fig:schematicview}(a)).
On the other hand, in magnetic-field much higher than 
$H_{\textrm{c1}}$, the penetration depth is defined to be a decay length of
magnetic-field from the center of a vortex core (Fig. \ref{Fig:schematicview}(b)).
They should fundamentally take the same value in an ideal system.
But, if a surface state different from the bulk one exists, the 
penetration depth obtained from low magnetic-field technique may deviate 
from that of the bulk state.
As an example, in $\beta$-(BEDT-TTF)$_2$PF$_6$, a structural relaxation in 
the surface BEDT-TTF molecules has been observed,\cite{362} as a characteristic 
feature of the surface state.
Besides, it is likely that only a small amount of crack or minor damage 
after quenching leads to a critical influence to the low-field 
measurements, which may give overestimation of $\lambda_{\parallel}(0)$.
The sample-independence in the data can check this point, 
but it seems not to be achieved in the literature.\cite{785}
This might cause a serious problem on the temperature dependence of 
the penetration depth as well, and thus a reconsideration will be needed for the 
experimental results performed at low magnetic-fields, especially after 
cooling fast.
Although the reports on $\lambda_{\parallel}(T)$ in the 
shielding state\cite{793,994,108,20,785} indicate the power-law 
behavior of $T^n$, it is still controversial whether $n$ = 1, 2, or 
between 1--2.
The discussion of the pairing symmetry from the $T$-dependence of 
$\lambda_{\parallel}$ may not become conclusive unless this 
problem on the absolute value of $\lambda_{\parallel}(0)$ is clarified.

\begin{figure}
\begin{center}
\includegraphics[clip]{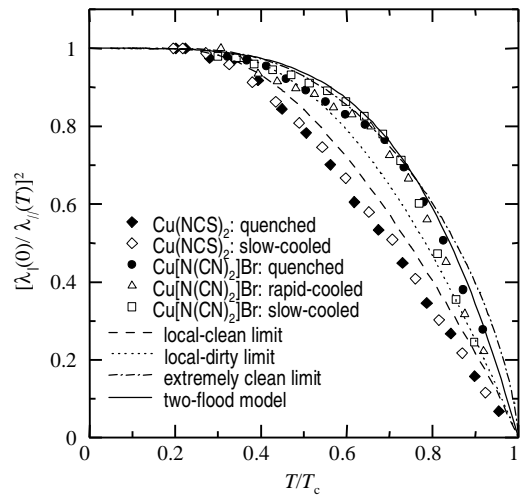}
 \caption{\label{Fig:rho} $[\lambda_{\parallel}(0)/ 
 \lambda_{\parallel}(T)]^2(\equiv\rho)$ as a  
 function of $T/T_{\textrm{c}}$. The four curves represent the models based on 
 the conventional BCS theory.\cite{Tinkham}
 }
\end{center}
\end{figure}

Next we discuss the insensitivity in $\lambda_{\parallel}(0)$ by 
quenching for the Cu(NCS)$_2$ salt.
In this salt, no change of $\lambda_{\parallel}(0)$ by quenching is
observed within the experimental accuracy
in contrast to the Cu[N(CN)$_2$]Br salt.
Nevertheless, $T_{\textrm{c}}$ in the Cu(NCS)$_2$ salt slightly decreases by quenching as 
well as the Cu[N(CN)$_2$]Br salt. 
This implies that the ethylene-disorders are surely introduced to the 
system after quenching in the Cu(NCS)$_2$ salt as well, and thus the mean free path 
will be also reduced.
In a high-quality sample, the mean free path in the slow-cooled state 
$l_\parallel\sim$ 200 nm is obtained from $T_{\textrm{D}}$ = 0.28 K.\cite{159}
As an example, assuming that $l_{\parallel}$ becomes half of this value by quenching,
one can estimate 
$\lambda^{\textrm{q}}_{\parallel}(0)/\lambda^{\textrm{s}}_{\parallel}(0)=1.01$ by using eq. (\ref{Eq:London}),
which is smaller than the experimental error (= $\pm$20 nm/430 nm $\sim\pm$5\%).
Thus the insensitivity in $\lambda_{\parallel}(0)$ is 
certainly reasonable as a very clean system with large $l_{\parallel}$ compared with the
Cu[N(CN)$_2$]Br salt.

Finally we comment on the temperature dependence of the penetration depth.
In our investigation, the cooling-rate seems not to change the form of 
$\lambda_{\parallel}(T)$ as shown in Fig. \ref{Fig:lambda}.
Figure \ref{Fig:rho} shows $[\lambda_{\parallel}(0)/\lambda_{\parallel}(T)]^2(\equiv\rho)$ 
as a function of $T/T_{\textrm{c}}$.
In each salt, $\rho(T/T_{\textrm{c}})$ is likely to be well scaled to the same 
functional form.
This indicates no influence of disorder's impurity on $\rho$ as a 
function of temperature, suggesting that the pairing symmetry is not altered.
Unfortunately, the correct estimate of $\lambda_{\parallel}(T)$ becomes 
difficult with lowering temperature in the present investigations, 
because $H_{\textrm{irr}}$ narrows the reversible magnetization region (shown in Figs. 
\ref{Fig:MH3}(c), \ref{Fig:MH3}(d), \ref{Fig:MH4}(c), and 
\ref{Fig:MH4}(d)) and the application of the London model becomes unfavorable.
This interrupts the accurate definition of $\lambda_{\parallel}$ 
at low temperatures.
We therefore do not conclude the pairing symmetry of the ground states.
Nevertheless, we can point out the difference between the two salts.
With increasing $T$, a faster decrease of $\rho(T/T_{\textrm{c}})$ in the 
Cu(NCS)$_2$ salt is observed than that in the Cu[N(CN)$_2$]Br salt.
In order to compare with the conventional BCS theory, 
several models\cite{Tinkham} are displayed in Fig. \ref{Fig:rho}. 
The data of the Cu(NCS)$_2$ salt seem to agree with the local-clean 
limit model, while those of the Cu[N(CN)$_2$]Br salt are located in 
between the two-flood and local-dirty limits.
These trends are consistent with the literatures,\cite{197,180} which may 
reflect the difference in the sample purity between the two salts.

In the above discussions, we adopted the conventional London model,
but it does not exclude the unconventional pairing symmetries of these materials.
The controversial problem on the functional form of 
$\lambda_{\parallel}(T)$ at low temperatures is still an open question.

\section{Conclusion}
In conclusion, we report the dc magnetization measurements 
under the control of cooling-rate for $\kappa$-(BEDT-TTF)$_2X$ ($X$ = 
Cu(NCS)$_2$ and Cu[N(CN)$_2$]Br).
The in-plane penetration depth is evaluated from the slope of the 
linear region in $M$ vs $\ln H$ plot.
We quantitatively explain the behavior of $\lambda_{\parallel}(0)$ 
by cooling fast for both salts in terms of the local-clean approximation.
In the Cu(NCS)$_2$ salt, the cooling-rate independent 
$\lambda_{\parallel}(0)$ = 430$\pm$20 nm is obtained within the 
experimental accuracy, while $T_{\textrm{c}}$ is slightly reduced.
This implies that this salt is described as a very clean system.
In contrast, a distinct increase of $\lambda_{\parallel}(0)$ in the 
Cu[N(CN)$_2$]Br salt is observed:
570$\pm30$ nm (slow-cooled), 590$\pm30$ nm (rapid-cooled), and 
690$\pm50$ nm (quenched).
This increase of the penetration depth is quantitatively in good 
agreement with the London model as shown in Fig. \ref{Fig:Londonfit}.
This demonstrates that the ethylene-disorders introduced by cooling 
faster increase the electron (quasiparticle)-scattering, resulting in 
shorter mean free path.

This research was partly supported by the Ministry of Education,
Science, Sports and Culture, Grant-in-Aid for Encouragement of Young
Scientists.


\begin{references}
\bibitem{estimate_l}The in-plane mean free path $l_{\parallel}$ is 
estimated from the Dingle temperature and effective mass obtained from 
the quantum oscillation measurements in Refs. \onlinecite{853,159,851,667,1025}.
In contrast, $l_{\parallel}$ obtained from the in-plane residual 
resistivity\cite{177,141} is much shorter than these values: 15--30 nm.
This discrepancy may be related to the domain structure as discussed in Ref. 
\onlinecite{1025}.
\end{references}
\end{document}